\documentclass[twocolumn,secnumarabic,amssymb, amsmath, nofootinbib,tightenlines, nobibnotes, aps, prl,epsfig]{revtex4}
\usepackage{graphicx}
\usepackage{dcolumn}
\usepackage{bm}
\begin{document}
\preprint{APS/123-QED}
\title{ The  color dipole model bounds with the gluon-gluon recombination correction}

\author{G.R.Boroun}%
 \email{grboroun@gmail.com; boroun@razi.ac.ir }
\author{B.Rezaei }
\altaffiliation{brezaei@razi.ac.ir}
\affiliation{ Physics Department, Razi University, Kermanshah
67149, Iran}

\date{\today}
\begin{abstract}
We present nonlinear (NL) and higher twist (HT) corrections to the
color dipole model (CDM) bounds at low values of $x$ and $Q^{2}$
using the parameterization method. Consistency between the bounds
at this region describe that a transition from the linear to the
nonlinear behavior is dependence on the behavior of the  gluon
distribution function. The parameters in the color dipole model
are comparable with the color dipole bounds at low values of
$Q^{2}$. Consequently, the obtained reduced cross sections at low
and moderate $Q^{2}$ values due to the NL+HT effects show a good
agreement with the H1
data.\\
\end{abstract}
 \pacs{***}
\maketitle
\subsection{1. Introduction}

The starting points on the color dipole model were given by
Sakurai and Schildknecht in 1972 [1] and  has been expanded so far
by some authors in Refs.[2,3,4]. The modern picture of the deep
inelastic scattering (DIS) at low $x$ is described as the color
dipole picture (CDP). In this picture the virtual photon
fluctuates into the $q\overline{q}$ pair  which this pair
interaction with the gluon field in the nucleon as a
gauge-invariant color-dipole interaction. Due to the interaction
of the  gluon fields with the $q\overline{q}$ dipole, the dipole
cross section, $\sigma_{(q\overline{q})^{J=1}_{L,T}p}$, is
described at the color transparency and saturation limits. The
$W^{2}$-dependent scale $\Lambda_{sat}^{2}(W^{2})$  separates the
two regions. The color transparency of the dipole cross section
according to the region of $Q^{2}{\gg}\Lambda_{sat}^{2}(W^{2})$
 and the saturation according to the region of
$Q^{2}{\ll}\Lambda_{sat}^{2}(W^{2})$ respectively. Indeed the
$(Q^{2},W^{2})$ plane of the CDP indicates that the line
$\eta(W^{2}, Q^{2}) = 1$ subdivides the $(Q^{2},W^{2})$  plane
into the saturation region of $\eta(W^{2}, Q^{2})< 1$ and the
color transparency region of $\eta(W^{2}, Q^{2})> 1$. $\eta(W^{2},
Q^{2})$ denotes the low-$x$ scaling variable, $\eta(W^{2},
Q^{2})=\frac{Q^{2}+m^{2}_{0}}{\Lambda_{sat}^{2}(W^{2})}$ which
$\Lambda_{sat}^{2}(W^{2})$ being the saturation scale and
$m_{0}^{2}{\simeq}0.15~\mathrm{GeV}^{2}$. At low-$x$ scaling, the
total photoabsorption cross section
$\sigma_{\gamma^{*}p}(W^{2},Q^{2})=\sigma_{\gamma^{*}p}(\eta(W^{2},Q^{2}))$
is described as $\log(1/\eta(W^{2},Q^{2}))$ for
$\eta(W^{2},Q^{2})<1$ and as $1/\eta(W^{2},Q^{2})$ for
$\eta(W^{2},Q^{2}){\gg}1$. At large
$Q^{2}{\gg}\Lambda_{sat}^{2}(W^{2})$, the
longitudinal-to-transverse ratio of the photoabsorption cross
sections $\sigma_{\gamma^{*}_{L}p}(W^{2},Q^{2})$ and
$\sigma_{\gamma^{*}_{T}p}(W^{2},Q^{2})$ reads as
\begin{eqnarray}
R(W^{2},Q^{2})=\frac{\sigma_{\gamma^{*}_{L}p}(W^{2},Q^{2})}{\sigma_{\gamma^{*}_{T}p}(W^{2},Q^{2})}
=\frac{1}{2\rho}.
\end{eqnarray}
In terms of the proton structure functions, the ratio of the
structure functions becomes
\begin{eqnarray}
\frac{F_{L}(W^{2},Q^{2})}{F_{2}(W^{2},Q^{2})} =\frac{1}{1+2\rho}.
\end{eqnarray}
The parameter $\rho$ is associated with the enhanced transverse
size of $q\overline{q}$ fluctuations in the CDM.  This parameter
is originating from transverse,
$\gamma^{*}_{T}{\rightarrow}q\overline{q}$, and longitudinal,
$\gamma^{*}_{L}{\rightarrow}q\overline{q}$, photons. Indeed the
$\rho$ parameter describes  the ratio of the average transverse
momenta
$\rho=\frac{<\overrightarrow{k}^{2}_{\bot}>_{L}}{<\overrightarrow{k}^{2}_{\bot}>_{T}}$.
It can also be related to the ratio of the effective transverse
sizes of the $(q\overline{q})^{J=1}_{L,T}$ states as
$\frac{<\overrightarrow{r}^{2}_{\bot}>_{L}}{<\overrightarrow{r}^{2}_{\bot}>_{T}}=\frac{1}{\rho}$.
The $\rho$ parameter is assumed to be proportional to the singlet
structure and gluon distribution functions in the large-$Q^{2}$
limit [5]
\begin{eqnarray}
\rho(x,Q^{2})=\frac{3\pi}{8\alpha_{s}(Q^{2})}\frac{F_{2}^{s}(x,Q^{2})}{G(x,Q^{2})}-\frac{1}{2},
\end{eqnarray}
where $F_{2}^{s}(x,Q^{2})=x\Sigma(x,Q^{2})$ and
$G(x,Q^{2})=xg(x,Q^{2})$.\\
In this paper we want to show that the behavior of the CDM bounds
at low and moderate $Q^{2}$ values are depends  on the gluon
density  behavior. In this case the bounds are obtained via the
nonlinear-DGLAP (Dokshitzer-Gribov-Lipatov-Altarelli-Parisi)
evolution. Studies along this line not only confirm HERA
investigations but also provide crucial benchmarks for further
investigations of the high-energy limit of QCD at the Electron-Ion
Collider (EIC) [6] and the large Hadron Electron Collider (LHeC)
[7,8]. The kinematic extension of the LHeC will allow us to
examine the non-linear dynamics at low $x$. The non-linear region
is approached when the reaction is mediated by multi-gluon
exchange. Indeed the growth of the gluon density is slowed down at
very small $x$ by gluon -gluon recombination process. The
kinematic coverage of the NC $e^{-}p$ scattering pseudodata at the
LHeC which indicate the nonlinear dynamics are defined  in the
region $x<0.01$ and $Q^{2}<700~\mathrm{GeV}^{2}$ [9,10]. At small
$x$ the effect of $\varpropto{\ln(1/x)}$ terms on the linear
evolution equations increases. So nonlinear interactions must be
applied. Indeed we need reliable LHeC predictions to understand
the low $x$ physics [11]. Since non-linear dynamics are known to
become sizable only at small-$x$, so the non-linear contribution
to the evolution equation [12] leads to an equation of the form
\begin{eqnarray}
\frac{\partial^{2}xg(x,Q^{2})}{\partial{\ln}(1/x)\partial{\ln}Q^{2}}=
\overline{\alpha}_{s}xg(x,Q^{2})-\frac{9}{16}\overline{\alpha}_{s}^{2}\pi^{2}
\frac{[xg(x,Q^{2})]^{2}}{\mathcal{R}^{2}Q^{2}},
\end{eqnarray}
where $\overline{\alpha}_{s}{\equiv}{\alpha}_{s}C_{A}/\pi$ and the
parameter $\mathcal{R}$ controls the strength of the nonlinearity.
The second nonlinear term in (4) is responsible for gluon
recombination. This term arises from perturbative QCD diagrams
which couple four gluons to two gluons. So that two gluon ladders
recombine into a single gluon ladder. It leads to saturation of
the gluon density at low $Q^{2}$ with decreasing $x$ [13]. The
gluon recombination is as important as gluon splitting which in
analysis some groups such as MRST2001 [14] and CETQ6M [15] in NLO
analysis considered. This implies that towards small values of $x$
and $Q^{2}$, the problem of negative gluon distribution in these
groups appears. Other non-linear equations such as Modified-DGLAP
(Md-DGLAP) [16], Balitsky-Kovchegov (BK) [17] and
Jalilian-Marian-McLerran-Weigert-Leonidov-Kovner (JIMWLK) [18]
equations have been derived and considered in the last years.
 Some another models, such as the
impact-parameter dependent saturation model (IP-Sat) [9] ,
developed a dipole model for DESY HERA which incorporates the
impact parameter distribution of the proton. It is a simple dipole
model that incorporates key features of the physics of gluon
saturation. This model for the dipole amplitude contains an
eikonalized gluon distribution which satisfies DGLAP evolution
while explicitly maintaining unitarity [19]. In Ref.[20] the
nonlinear evolution equation for dipole density have been
developed. The deeply inelastic scattering at very high energies
in the saturation regime considered.\\
The unitarity problem is discussed in Ref.[21] with respect to
photoabsorption cross sections. The unitarity relation entails the
nonlinearity of the observed DIS structure functions in terms of
the impulse approximation (IA) parton densities. The expectation
value of the interaction cross sections of the multiparton Fock
states of the virtual photon over the wave functions is considered
in [21]. The unitarized total cross sections $\sigma (x,\rho)$
reads
$$\sigma (x,\rho)\simeq \sigma_{0}(x,\rho)~ \mathrm{at}~ \eta(x,\rho)\ll
1$$ where the quantity $\eta(x,\rho)$ controls the effect of the
unitarization and $\rho$ is the transverse size of the
$q\overline{q}$ pair. At $\eta(x,\rho)\gg 1$ the unitarization
suppresses the cross section as $\sigma (x,\rho)\ll
\sigma_{0}(x,\rho)$ where $\sigma_{0}(x,\rho)$ is the interaction
cross section for the $q\overline{q}$ color dipole of size $\rho$.
The effects of the $q\overline{q}g$ Fock state is deriving term of
the triple-pomeron mass spectrum. The shadowing term in the
unitarized structure function is dominated by the triple-pomeron
term, which is approximately independent of the flavor and $Q^{2}$
variables. Indeed the unitarity (shadowing) correction is a
nonlinear functional of the DGLAP cross section. Also the
unitarity correction can be related to the cross section of the
forward diffractive dissociation of the virtual photons (DDIS)
$\gamma^{\ast}+p{\rightarrow}X+p$ where $X=q\overline{q}$. The
conventional description of DDIS is based on the leading twist
DGLAP evolution equations which characterize the QCD hard scale
dependence of the diffractive parton distribution functions
(DPDFs) [22]. The effects of pomeron loops and running coupling on
the cross sections for inclusive $\gamma^{\ast}h$ and on
diffractive deep inelastic scattering are investigated in
Ref.[23]. In Ref.[24] DDIS provides  a basis for the definition of
the Weizs$\mathrm{\ddot{a}}$cker-Williams (WW) nuclear gluon
structure function. Also the initial conditions at low $x$ DIS off
nucleons and nuclei for QCD evolution that satisfy unitarity  are
described. The nonlinear effects can be tested at a
superior statistical accuracy attainable at EIC.\\
 This paper is organized as follows. In the next section the
 theoretical formalism is presented, including the nonlinear
 evolution and the  color dipole parameters.  In section 3, we present
  a detailed numerical analysis and our main results. We then confront these
results with  the CDM bounds at low values of $Q^{2}$. In the last
section  we summarize our main conclusions and remarks.\\

\subsection{2. Theoretical formalism}

In the CDM the $\rho$ parameter is dependent on the proton
structure function $F_{2}(x,Q^{2})$ and the gluon distribution
function $G(x,Q^{2})$ as reads
\begin{eqnarray}
\rho(x,Q^{2})=\frac{27\pi}{20\alpha_{s}(Q^{2})}\frac{F_{2}(x,Q^{2})}{G(x,Q^{2})}-\frac{1}{2}.
\end{eqnarray}
An analytical expression for $F_{2}(x,Q^{2})$ has suggested which
describes fairly well the available experimental data on the
reduced cross section [25]. This parameterization provides
reliable structure function $F_{2}(x,Q^{2})$ according to HERA
data at low $x$ in a wide range of the momentum transfer
$(1~\mathrm{GeV}^{2}<Q^{2}<3000~\mathrm{GeV}^{2})$ as
\begin{eqnarray}
F_{ 2}(x,Q^{2})& =& D(Q^{2})(1-
x)^{n}\sum_{m=0}^{2}A_{m}(Q^{2})L^{m},
\end{eqnarray}
and can be applied as well in analyses of ultra-high energy
processes with cosmic neutrinos. In a new method, the linear
behavior of the gluon density in the CDM parameters is
investigated in Ref.[26]. Now we consider the non-linear behavior
of the gluon density for the CDM bounds. The nonlinear effects of
the gluon-gluon fusion due to the high gluon density at small $x$
is considered in Gribov-Levin-Ryskin-Mueller-Qiu (GLR-MQ) [12].
Some studies of the GLR-MQ equation in the framework of the
extracting the gluon distribution function have been discussed
considerably over the past years [27-33]. The GLR-MQ equation can
be written in standard form [34]
\begin{eqnarray}
\frac{\partial{G(x,Q^{2})}}{\partial{\ln}Q^{2}}&=&\frac{\partial{G(x,Q^{2})}}{\partial{\ln}Q^{2}}|_{DGLAP}\nonumber\\
&&-\frac{81}{16}\frac{\alpha^{2}_{s}(Q^{2})}{\mathcal{R}^{2}Q^{2}}\int_{\chi}^{1}\frac{dz}{z}G^{2}(\frac{x}{z},Q^{2}),
\end{eqnarray}
where $\chi=\frac{x}{x_{0}}$ and $x_{0}$ is the boundary condition
that the gluon distribution joints smoothly onto the linear
region. The correlation length $\mathcal{R}$  determines the size
of the nonlinear terms. This value depends on how the gluon
ladders are coupled to the nucleon or on how the gluons are
distributed within the nucleon. The $\mathcal{R}$ is approximately
equal to $\simeq 5~\mathrm{GeV}^{-1}$ if the gluons are populated
across the proton and it is equal to $\simeq 2~\mathrm{GeV}^{-1}$
if the gluons have hotspot like structure. By solving GLR-MQ
(Eq.7), we obtain an expression for the nonlinear gluon
distribution function (i.e., $G^{NL}(x,Q^{2})$ ) as
\begin{eqnarray}
G^{\mathrm{NL}}(x,Q^{2})&=&G^{\mathrm{NL}}(x,Q_{0}^{2})+G(x,Q^{2})-G(x,Q_{0}^{2})\\
&&-\int_{Q_{0}^{2}}^{Q^{2}}\frac{81}{16}\frac{\alpha^{2}_{s}(Q^{2})}{\mathcal{R}^{2}Q^{2}}\int_{\chi}^{1}\frac{dz}{z}G^{2}(\frac{x}{z},Q^{2})d{\ln}Q^{2}\nonumber
\end{eqnarray}
We note that at $x{\geq}x_{0}(=10^{-2})$ the linear and nonlinear
gluon distribution behaviors are equal. At $Q_{0}^{2}$ the low $x$
behavior of the nonlinear gluon distribution is assumed to be [35]
\begin{eqnarray}
G^{\mathrm{NL}}(x,Q_{0}^{2})&=&G(x,Q_{0}^{2})\{1+\frac{27\pi{\alpha_{s}(Q_{0}^{2})}}{16\mathcal{R}^{2}Q_{0}^{2}}\theta(x_{0}-x)\nonumber\\
&&{\times}[G(x,Q_{0}^{2})-G(x_{0},Q_{0}^{2})] \}^{-1}.
\end{eqnarray}
Substituting Eqs.(6) and (8) in Eq.(5) the nonlinear behavior of
the $\rho$ parameter becomes
\begin{eqnarray}
\rho^{\mathrm{NL}}(x,Q^{2})=\frac{27\pi}{20\alpha_{s}(Q^{2})}\frac{F_{2}(x,Q^{2})(i.e.,Eq.(6))}{G^{\mathrm{NL}}(x,Q^{2})(i.e.,Eq.8)}-\frac{1}{2}
\end{eqnarray}
Next we define the nonlinear behavior of the
longitudinal-to-transverse cross sections and the structure
functions by the following forms respectively
\begin{eqnarray}
R^{\mathrm{NL}}(W^{2},Q^{2})=\frac{1}{2\rho^{\mathrm{NL}}(W^{2},Q^{2})},
\end{eqnarray}
and
\begin{eqnarray}
F_{L/2}^{\mathrm{NL}}(W^{2},Q^{2}){\equiv}\frac{F_{L}(W^{2},Q^{2})}{F_{2}(W^{2},Q^{2})}
=\frac{1}{1+2\rho^{\mathrm{NL}}(W^{2},Q^{2})}
\end{eqnarray}
If we rewrite the reduced cross section in terms of the nonlinear
behavior of the ratio of the structure functions, then the
nonlinear behavior of the reduced cross section at low $Q^{2}$
reads
\begin{eqnarray}
\sigma_{r}^{\mathrm{NL}}(W^{2},Q^{2})&=&F_{2}(W^{2},Q^{2})[1-\frac{y^{2}}{1+(1-y)^2}\nonumber\\
&&{\times}\frac{1}{1+2\rho^{\mathrm{NL}}(W^{2},Q^{2})}].
\end{eqnarray}
Here $W^{2}{\simeq}sy$ which the inelasticity $y$ is related to
$Q^{2}$, $x$ and the center-of-mass energy squared,
$s=4E_{e}E_{p}$, by $y=Q^{2}/sx$.\\
In the following we consider the deeply inelastic structure
functions at low $Q^{2}$ using the higher-twist (HT) corrections
in QCD.  Using this effect in the parameterization of the proton
structure function is expected to provide better results for the
reduced cross section than the experimental data. The higher-twist
corrections arise from the struck proton$^{,}$s interaction with
target remnants where reflecting confinement [36-40]. The
phenomenological power correction to the structure function from
the HT corrections
 is considered by the following form
 \begin{eqnarray}
F_{2}^{\mathrm{HT}}(x,Q^{2})=F_{2}^{\mathrm{Parameterization}}(x,Q^{2})(1+\frac{C_{HT}(x)}{Q^{2}})
 \end{eqnarray}
which the coefficient function $C_{HT}(x)$ is determined from fit
to the data. In some references [37-41] this quantity is set to be
an free parameter as $C_{HT}=0.12~\pm~0.07~\mathrm{GeV}^{2}$ and
in others [42,43] it depends on $x$ as
\begin{eqnarray}
C_{HT}(x)=h_{0}(h_{2}(x)x^{h_{1}}+\gamma).\nonumber
\end{eqnarray}
In Refs.[42,43] this fit parameterization is obtained  from the QCD analysis with the HT corrections included. \\
Therefore it is clear from Eqs.(13) and (14) that at low $Q^{2}$,
we can add the HT corrections and our solution takes the form
\begin{eqnarray}
\sigma_{r}^{\mathrm{NL+HT}}(W^{2},Q^{2})&=&F_{2}^{\mathrm{HT}}(W^{2},Q^{2})[1-\frac{y^{2}}{1+(1-y)^2}\nonumber\\
&&{\times}\frac{1}{1+2\rho^{\mathrm{NL+HT}}(W^{2},Q^{2})}],
\end{eqnarray}
with
\begin{eqnarray}
\rho^{\mathrm{NL+HT}}(x,Q^{2})=\frac{27\pi}{20\alpha_{s}(Q^{2})}\frac{F_{2}^{\mathrm{HT}}(x,Q^{2})}{G^{\mathrm{NL}}(x,Q^{2})}-\frac{1}{2}.
\end{eqnarray}

\subsection{3. Results and discussions}
In this paper, we obtain the nonlinear gluon distribution function
solving the GLR-MQ evolution equation for gluon density. The
analysis is performed in the range $10^{-5}{\leq}x{\leq}10^{-2}$
and $1{\leq}Q^{2}{\leq}100~\mathrm{GeV}^{2}$. The computed results
of nonlinear gluon distribution function are compared with the CDP
model [5] (Kuroda and Schildknecht, Phys.Rev.D85, 094001(2011))
and the parameterization model [25] in Fig.1. According to Fig.7
in Ref.[5] (Kuroda and Schildknecht, Phys.Rev.D85, 094001(2011)),
there is a considerable agreement to the results from the CETQ
[44] and  MSTW [45] collaborations. The nonlinear gluon
distribution behavior is comparable with MSTW08 NNLO [46]
at $Q^{2}>1~\mathrm{GeV}^{2}$.\\
In the following, the parameters and bounds with respect to the
nonlinear gluon distribution behavior can be examined. With the
obtained $\rho$ parameter, we calculate the ratio of structure
functions and also the reduced cross sections with respect to the
nonlinear and higher twist corrections. These functions are
obtained at low $x$ and $Q^{2}$ values by taking an appropriate
input parton distribution. In Fig.2, the parameters $\rho$, $R$
and $F_{L/2}$ are obtained with respect to the nonlinear behavior
of the gluon distribution function. In the following we have
investigated the effect of nonlinearity in our results in the
hot-spot point. The value of this parameter is defined to be
$R=2~\mathrm{GeV}^{-1}$ in this work. In Fig.2 we shown that the
nonlinear results are much closer to the color dipole bounds than
the linear ones. The comparison is for $Q^{2}=5~\mathrm{GeV}^{2}$
and $Q^{2}=10~\mathrm{GeV}^{2}$. The fluctuations corresponding to
the parameters (i.e., $\rho$, $R$ and $F_{L/2}$) in comparison
with constant CDM bounds are due to the parameterization of the
PDFs. By adding the effect of the HT corrections  on the
parameters, we showed that the results have a behavior comparable
to the CDM bounds. In Fig.3, a comparison for
$Q^{2}=5~\mathrm{GeV}^{2}$ has been made between the nonlinear and
nonlinear+higer twist (NL+HT) corrections to the parameters. In
the following we will apply these corrections (i.e., NL+HT) to all
results. As can be observe in Fig.4, the ratio of the structure
functions are comparable to the H1 data [47] and CDM bounds
[3,4,48] not only at large $Q^{2}$ but also at low $Q^{2}$ values.
Indeed, the transition from the linear to nonlinear is done due to
the nonlinear corrections to the gluon distribution function.
Compared to other results and models, we see that the ratio
$F_{L}/F_{2}$ is in fact comparable to the results of others
[10,49]  and experimental data. This comparison is very good at
low and high-$Q^{2}$ values, even compared to other models such as
Golec-Biernat-W$\mathrm{\ddot{u}}$sthoff (GBW)[10] and
Iancu-Itakura-Munier (IIM) [49] parameterizations. The nonlinear
behavior of the ratio of structure functions at low $Q^{2}$ in
Fig.5 is observable in comparison with the H1 data [41]. In Fig.5,
data collected in the region of low momentum transfers,
$0.2~\mathrm{GeV}^{2}{\leq}Q^{2}{\leq}12~\mathrm{GeV}^{2}$, and
low Bjorken $x$, $10^{-6}{\lesssim}~x~{\lesssim}0.02$ with
center-of-mass energy $\sqrt{s}=319~\mathrm{GeV}$. In Ref.[41] the
structure functions of $F_{2}$ and $F_{L}$  collected without the
total errors. Tables 17 and 18 in this reference shown that
$F^{th}_{L}$ represents the structure function $F_{L}$ used for
the center-of-mass energy (CME) correction and to calculate the
structure function $F_{2}$. Therefore we compared our results at
$x{=}0.001$  in a wide range of $W^{2}$ between
$10^{3}~\mathrm{GeV}^{2}$ and $10^{4}~\mathrm{GeV}^{2}$ with the
ratio of structure functions (i.e., H1 2009 [41]) without the
total
uncertainties in Fig.5.\\
In the following we use the NL+HT behavior of the ratio
$F_{L}/F_{2}$ to calculate the reduced cross section. In Ref.[41]
the H1 collaboration  reported the DIS cross sections at low
$Q^{2}$. The DIS data  collected  based on the SVX, NVX-BST and
NVX-S9 analysis [41]. We use the SVX data at
$Q^{2}=2~\mathrm{GeV}^{2}$, the NVX-BST and NVX-S9 data at
$Q^{2}=5~\mathrm{GeV}^{2}$ and the NVX-BST
 data at $Q^{2}=12~\mathrm{GeV}^{2}$. The cross section data due
 to the NL+HT effects at three values  of $Q^{2}$ are given in
 Table I, and compared with the H1 data [41] measured from the SVX and NVX
 data. Here we discuss the $\chi^{2}$ method for comparison
  according to the number of points at any $Q^{2}$ values. The $\chi^{2}$ can be defined as
\begin{eqnarray}
\chi^{2}=\Huge{\sum_{i}^{N_{\mathrm{data}}}}(X_{\mathrm{data},i}
-X_{\mathrm{method},i})^{2}/(\delta{X}_{i})^{2},
\end{eqnarray}
where $i$ runs all the data points, $\delta{X}_{i}$ can be the
total experimental uncertainties. The $\chi^{2}/N_{data}$ can
quantify the agreement between the data and our predictions. The
$\chi^{2}/N_{data}$ computed at different values of $Q^{2}$ is in
Table II. The results are plotted in Fig.6. In this figure,
together with the H1 data [41], we plot the
 $x$ dependence of the reduced cross section $\sigma_{r}(x,Q^{2})$
 computed with respect to the nonlinear and higher-twist effects for fixed values of
 $Q^{2}$. The error bands are in
accordance with the statistical errors of the parameterization of
$F_{2}$ and the errors bares are quoted in $\%$ relative to
$\sigma_{r}$. The agreement of these results with the H1 data are
excellent for $x<0.01$. We also compare these results with  the
HERA data [50] (which combines H1 and ZEUS data) in Fig.6 at
$Q^{2}$ values of $2$ and $12~\mathrm{GeV}^{2}$. This includes
data taken with proton beam energies of $E_{p}=920~\mathrm{GeV}$
corresponding to the center-of-mass energy $\sqrt{s}=
318~\mathrm{GeV}$. Therefore, the results of the current paper in
the region of smallest $x$ and $Q^{2}$ studies can confirm the
nonlinear corrections to the small-$x$ gluon distributions for
transition.\\

\subsection{4. Summary}
In conclusion, we have studied the effects of adding the nonlinear
corrections to the gluon density for transition from the linear to
the nonlinear regions. We use the parameterization of
$F_{2}(x,Q^{2})$ as a baseline. This analysis is also enriched
with the higher twist (HT) contributions to the proton structure
function at small values of $Q^{2}$. The nonlinear and higher
twist corrections to the ratio of structure functions and also to
the reduced cross sections are considered. Comparing these
parameters with the CDM bounds indicate that the NL+HT effects are
enriched the behavior at low $Q^{2}$. The transition of the ratio
$F_{L/2}$ from the linear to the nonlinear behavior is considered
and shown that it is in good agreement with the CDM bounds non
only at high-$Q^{2}$ but also at low-$Q^{2}$ values. Comparison of
the reduced cross sections with respect to the nonlinear and
higher twist corrections with HERA data at low and moderate
$Q^{2}$ values shows that
this transition has been done with good accuracy in comparison with the HERA data. \\


\subsection{ACKNOWLEDGMENTS}

Authors are grateful the Razi University for financial support of
this project. G.R.Boroun is especially grateful to D.Schildknecht
for carefully
reading the manuscript and fruitful discussions.\\


\newpage
\section{References}
1. J.J.Sakurai and D.Schildknecht, Phys.Lett.B{\bf40}, 121(1972);
B.Gorczyca and D.Schildknecht, Phys.Lett.B{\bf47},
71(1973).\\
2. N.N.Nikolaev and B.G.Zakharov, Z.Phys.C{\bf49}, 607(1991);
Z.Phys.C{\bf53}, 331(1992); A.H.Mueller, Nucl.Phys.B{\bf415},
373(1994); K.Golec-Biernat and M.W$\mathrm{\ddot{u}}$sthoff,
Phys.Rev.D{\bf59}, 014017(1998);  H.Kowalski, L.Motyka and
G.Watt, Phys.Rev.D{\bf74}, 074016(2006); B.Sambasivam, T.Toll and T.Ullrich;
 Phys.Lett.B{\bf803}, 135277(2020); G.M.Peccini, F.Kopp, M.V.T.Machado and D.A.Fagundes,
 Phys.Rev.D{\bf101}, 074042 (2020).\\
3. M.Kuroda and D.Schildknecht, Phys.Lett. B{\bf618}, 84(2005);
 M.Kuroda and D.Schildknecht, Phys.Rev. D{\bf96}, 094013(2017);
 D.Schildknecht, B.Surrow and M.Tentynkov,
 Eur.Phys.J.C{\bf20}, 77(2001); M.Kuroda and D.Schildknecht, International Journal of Modern Physics A{\bf31}, No. 30, 1650157 (2016).\\
4. D.Schildknecht, Nuclear Physics B Proceedings Supplement 146,
211 (2005) ; Nuclear Physics B Proceedings Supplement 00, 1
(2012); F.Schrempp and A.Utermann, Acta Phys.Polon.B{\bf33},
3633(2002).\\
5. M.Kuroda and D.Schildknecht, Phys.Lett. B{\bf670}, 129(2008);
M.Kuroda and D.Schildknecht, Phys.Rev.D{\bf85}, 094001(2011).\\
6. D.Boer et al., arXiv: [nucl-th]1108.1713.\\
7. J.Abelleira Fernandez et al., [LHeC Study Group Collaboration],
J.Phys.G{\bf39}, 075001(2012); A. Abada et al., [FCC Study Group Collaboration], Eur.Phys.J.C{\bf 79}, 474(2019).\\
8. P.Agostini et al. [LHeC Collaboration and FCC-he Study Group ],
 CERN-ACC-Note-2020-0002, arXiv:2007.14491 [hep-ex] (2020).\\
9. H.Kowalski and D.Teaney, Phys.Rev.D{\bf68}, 114005(2003);
G.Watt and H.Kowalski,  Phys.Rev.D{\bf78}, 014016(2008).\\
10. J.Bartels, K.Golec-Biernat and H.Kowalski,  Phys.Rev.D{\bf66},
014001(2002); K.Golec-Biernat and M.W$\mathrm{\ddot{u}}$sthoff,
Phys.Rev.D{\bf60}, 114023(1999); K.Golec-Biernat and S.Sapeta, J.High Energ.Phys.{\bf03}, 102(2018). \\
11. M.Klein,  arXiv [hep-ph]:1802.04317; M.Klein,
Ann.Phys.{\bf528}, 138(2016); N.Armesto et al.,
Phys.Rev.D{\bf100}, 074022(2019).\\
12. L.V.Gribov, E.M.Levin and M.G.Ryskin, Phys.Rept.{\bf100},
1(1983); A.H.Mueller and J.w.Qiu, Nucl.Phys.B{\bf268},
427(1986).\\
13. M.R.Pelicer et al., Eur.Phys.J.C{\bf79}, 9(2019).\\
14. A.D.Martin, R.G.Roberts, W.J.Stirling and R.S.Thorne,
Eur.Phys.J.C{\bf23}, 73(2002); Phys.Lett.B{\bf531}, 216(2002).\\
15. J.Pumplin et al., J.High Energ.Phys.{\bf07}, 012(2002).\\
16. W.Zhu, J.Ruan, J.Yang and Z.Shen, Phys.rev.D{\bf68},
094015(2003).\\
17. I.Balitsky, Nucl.Phys.B{\bf463}, 99(1996); Y.V.Kovchegov,
phys.Rev.D{\bf60}, 034008(1999).\\
18. J.Jalilian-Marian, A.Kovner, a.Leonidov and H.Weigert,
Nucl.Phys.B{\bf504}, 415(1997); Phys.Rev.D{\bf59}, 014014(1998);
E.Iancu, A.Leonidov and L.D.McLerran, Nucl.Phys.A{\bf692},
583(2001); Nucl.Phys.A{\bf703}, 489(2002).\\
19. A.H.Rezaeian et al., Phys.Rev.D{87}, 034002(2013).\\
20. I.I.Balitsky and A.V.Belitsky, Nucl.Phys.B{\bf629},
290(2002).\\
21. V. Barone et al., Phys.Lett.B{\bf326}, 161(1994).\\
22. M. Sadzikowski, L. Motyka and  W. Slominski, arXiv
[hep-ph]:1206.1732 (2012).\\
23. M.B.Gay Ducati, E.G.de Oliveira and J.T.de Santana Amaral,
arXiv [hep-ph]:1209.5354 (2012).\\
24. I.P.Ivanov et al., XXXII International Symposium on
Multiparticle Dynamics, pp. 169-176 (2003);
 K.Tywoniuk, Journal of Physics: Conference Series{\bf270}, 012054 (2011).\\
25. M. M. Block, L. Durand and P. Ha, Phys. Rev.D{\bf 89}, no. 9,
094027 (2014).\\
26. G.R. Boroun, arXiv:2102.04867 (2021).\\
27. M.Devee and J.K.Sarma, Eur.Phys.J.C{74}, 2751(2014);
Nucl.Phys.B{\bf885}, 571(2014); P.Phukan, M.Lalung and J.K.Sarma,
Nucl.Phys.A{\bf968}, 275(2017); M.Lalung, P.Phukan and J.K.Sarma,
Nucl.Phys.A {\bf992}, 121615(2019); M.Devee, arXiv:1808.00899
[hep-ph] (2018);
 P.Gilvana and S.K.Werner, arXiv:1804.10659 [hep-ph] (2018).\\
28. G.R.Boroun, Eur.Phys.J.A{\bf42}, 251(2009).\\
29. G.R.Boroun, Phys.Rev.C{\bf97}, 015206(2018).\\
30. B.Rezaei and G.R.Boroun, Phys.Rev.C{\bf101}, 045202(2020).\\
31. B.Rezaei and G.R.Boroun, Eur.Phys.J.A{\bf56}, 262(2020).\\
32. G.R.Boroun and
S.Zarrin, Eur.Phys.J.Plus{\bf128}, 119(2013).\\
33. G.R.Boroun and B.Rezaei, Nucl.Phys.A{\bf1006}, 122062 (2021).\\
34. K.Prytz, Eur.Phys.J.C{\bf22}, 317(2001); K.J.Eskola et al.,
Nucl.Phys.B{\bf660}, 211(2003); M.A Kimber, J.Kwiecinski and
A.D.Martin, Phys.Lett.B{\bf508}, 58(2001).\\
35. A.D.Martin et al., Phys.Rev.D{\bf47}, 867(1993); J.Kwiecinski,
A.D.Martin and P.J.Sutton, Phys.Rev.D{\bf44}, 2640(1991);
A.J.Askew, J.Kwiecinski,
A.D.Martin and P.J.Sutton, Phys.Rev.D{\bf47}, 3775(1993).\\
36. B.Badelek et al., J.Phys.G\textbf{22}, 815(1996); S. Catani
and F. Hautmann, Nucl.Phys.B\textbf{427}, 475(1994); J.Blumlein
and H.Bottcher, arXiv[hep-ph]:0807.0248(2008).\\
37. A.M.Cooper-Sarkar et al., arXiv:1605.08577v1 [hep-ph] 27 May
2016.\\
38. I.Abt et.al., Phys.Rev.D{\bf94}, 034032(2016).\\
39. G.R.Boroun and B.Rezaei,  Phys.Lett.B{\bf816}, 136274 (2021).\\
40. G.R.Boroun and B.Rezaei, Nucl.Phys.A{\bf990}, 244(2019).\\
41. F.D. Aaron et al. [H1 Collaboration], Eur.Phys.J. C{\bf63},
625(2009).\\
42. H.Khanpour, A.Mirjalili and S.Atashbar Tehrani,
Phys.Rev.C{\bf95}, 035201 (2017).\\
43. H.Khanpour, Phys.Rev.D{\bf99}, 054007(2019).\\
44. J. Pumplin et al., [CTEQ Collaboration] JHEP{\bf0207}, 012(2002).\\
45. A.D. Martin et al., Eur. Phys. J. C{\bf18}, 117(2000).\\
46. A.D. Martin et al., Eur. Phys. J. C{\bf63}, 189(2009).\\
47. V.Andreev et al. [H1 Collaboration], Eur.Phys.J.C{\bf74},
2814(2014).\\
48. C.Ewerz, A.von Manteuffel and O.Nachtmann, JHEP{\bf03},
102(2010); D.Britzger et al., Phys.Rev.D{\bf100}, 114007 (2019);
C.Ewerz, A. von Manteuffel
 and O.Nachtmann, Phys.Rev.D{\bf77}, 074022(2008); C.Ewerz and O.Nachtmann, Phys.Lett.B{\bf648}, 279 (2007).\\
49. E.Iancu,K.Itakura and S.Munier, Phys.Lett.B{\bf590},
199(2004); M.Niedziela and M.Praszalowicz, Acta Physica Polonica
B{\bf46}, 2018(2015).\\
50. H.Abramowicz et al., [H1 and ZEUS Collaborations],
Eur.Phys.J.C{\bf75}, 580(2015).\\
\begin{figure}
\includegraphics[width=1\textwidth]{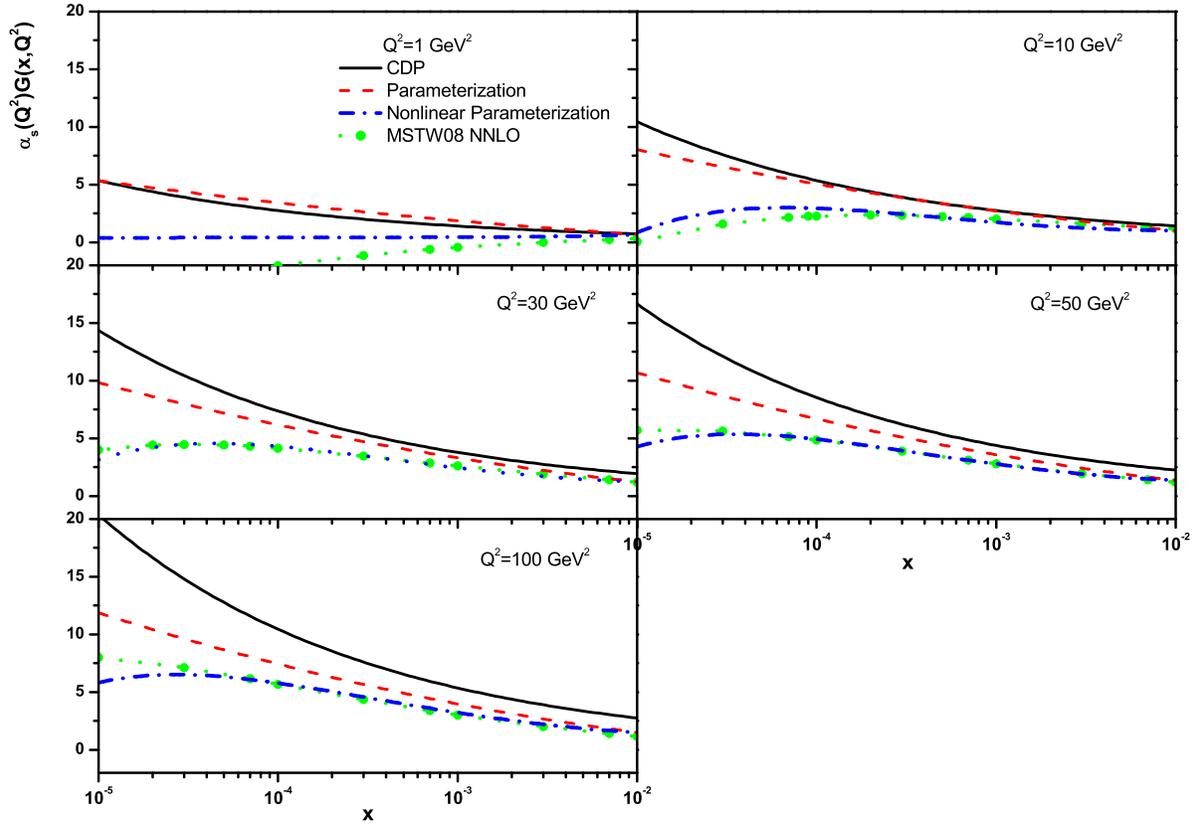}
\caption{The nonlinear gluon distribution function at
$R=2~\mathrm{GeV}^{-1}$ compared with the gluon distributions from
the CDP model [5] (Kuroda and Schildknecht, Phys.Rev.D85,
094001(2011)), the parameterization model [25] and MSTW08 NNLO
[46].}\label{Fig1}
\end{figure}
\begin{figure}
\includegraphics[width=0.55\textwidth]{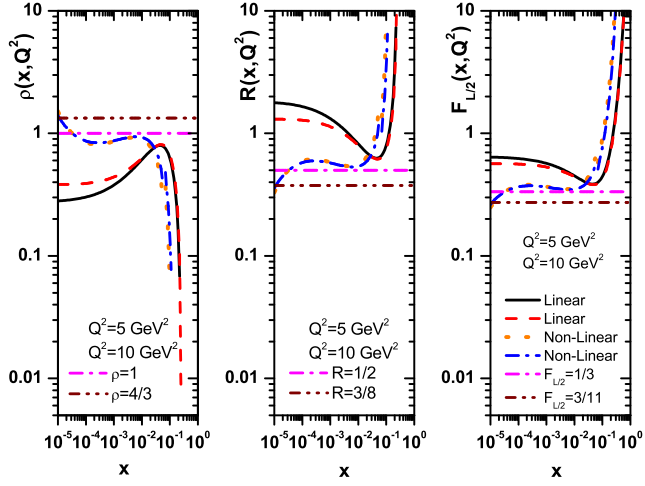}
\caption{Results of the parameters $(a):~\rho(x,Q^{2}),~
(b):~R(x,Q^{2})$ and $(c):~F_{L/2}(x,Q^{2})$ obtained from the
linear and nonlinear corrections at fixed $Q^{2}$ values
($Q^{2}=5~ \mathrm{GeV}^{2}$,Black-linear and Orange-nonlinear;
$Q^{2}=10~ \mathrm{GeV}^{2}$,Red-linear and Blue-nonlinear )
 respectively. The parameters
compared with the CDP bounds in $(a):$ $\rho=1$ and $4/3$, in
$(b):$ $R=1/2$ and $3/8$ and in $(c):$ $ F_{L/2}=1/3$ and $3/11$
respectively.}\label{Fig1}
\end{figure}
\begin{figure}
\includegraphics[width=0.55\textwidth]{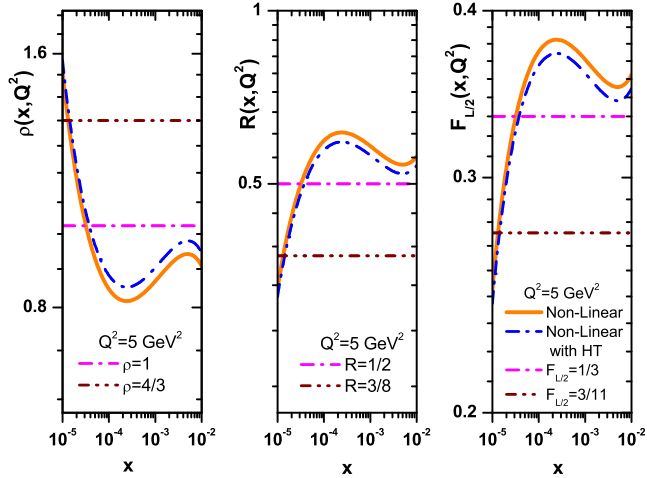}
\caption{Comparison of the  nonlinear behavior of the parameters
$(a):~\rho(x,Q^{2}),~ (b):~R(x,Q^{2})$ and $(c):~F_{L/2}(x,Q^{2})$
with the higher-twist corrections at $Q^{2}=5~\mathrm{GeV}^{2}$.
The parameters compared with the CDP bounds in $(a):$ $\rho=1$ and
$4/3$, in $(b):$ $R=1/2$ and $3/8$ and in $(c):$ $ F_{L/2}=1/3$
and $3/11$ respectively.}\label{Fig1}
\end{figure}
\begin{figure}
\includegraphics[width=0.65\textwidth]{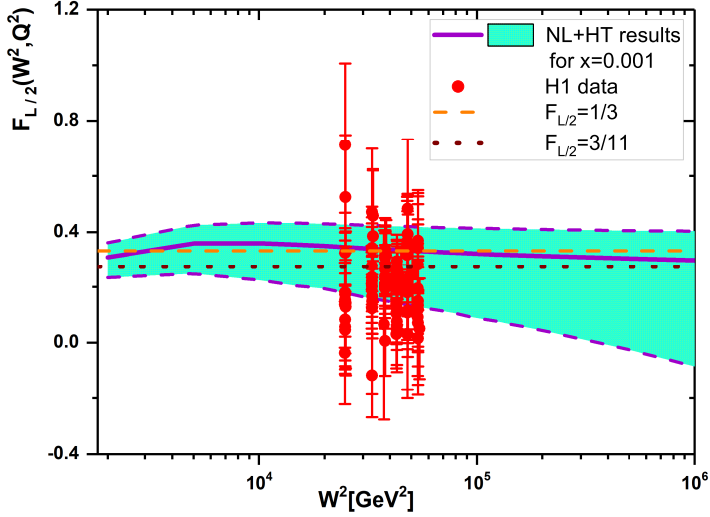}
\caption{The ratio of the longitudinal to transversal structure
functions calculated due to the nonlinear and higher twist effects
at fixed value of the Bjorken variable $x=0.001$. Experimental
data are from the H1 Collaboration as accompanied with total
errors [47]. The obtained values compared with the CDP bounds
[3,4,48] $F_{L/2}=1/3$ and $3/11$. The error bands are due  to the
effective parameters in the parameterization of $F_{2}(x,Q^{2})$
[25] and also the HT coefficient errors
[36-40,42,43].}\label{Fig1}
\end{figure}
\begin{figure}
\includegraphics[width=0.55\textwidth]{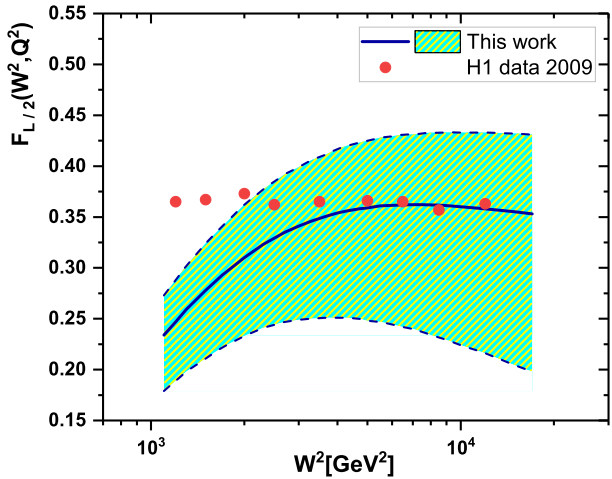}
\caption{ Continue Fig.3 in the low $Q^{2}$ values. The ratio of
the longitudinal to transversal structure functions calculated due
to the nonlinear and higher twist effects at fixed value of the
Bjorken variable $x=0.001$. Data are from the H1 Collaboration
[41] without the total uncertainties at low $Q^{2}$. The error
bands are due to the effective parameters in the parameterization
of $F_{2}(x,Q^{2})$ [25] and also the HT coefficient errors
[36-40,42,43].}\label{Fig1}
\end{figure}
\begin{figure}
\includegraphics[width=1\textwidth]{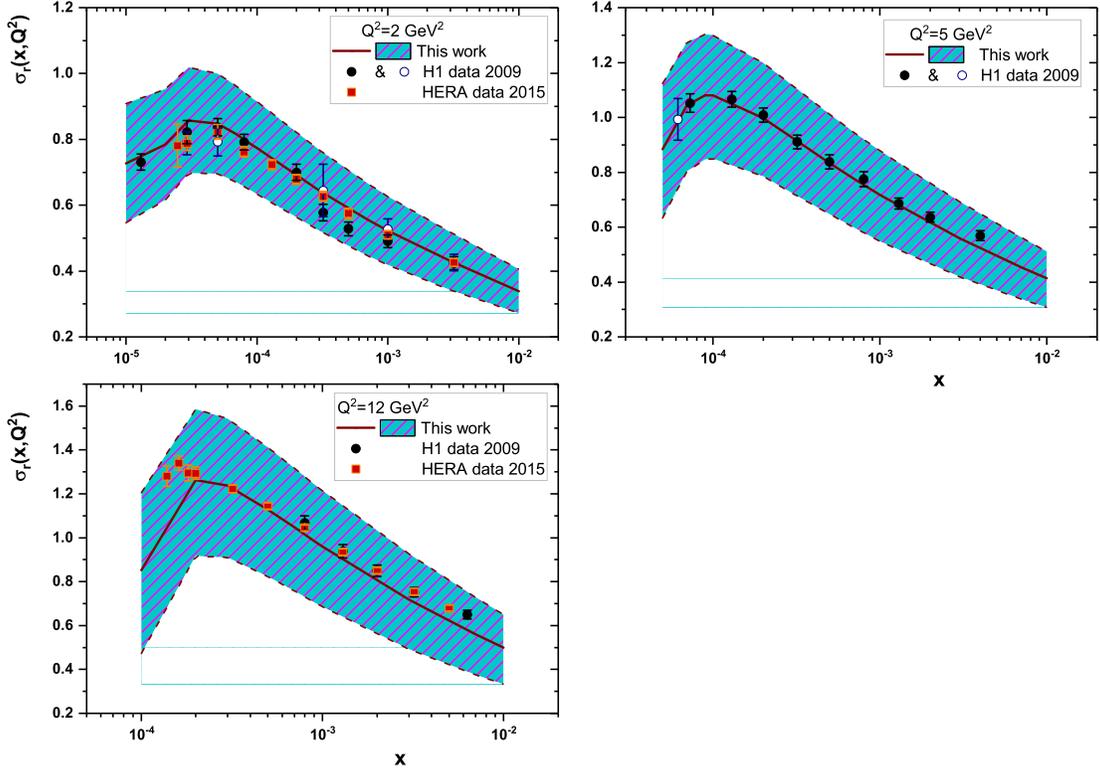}
\caption{Reduced cross section $\sigma^{NL+HT}_{r}$ from the
nonlinear behavior of the gluon distribution and the higher twist
corrections to the proton structure function at low $x$ and
$Q^{2}$ compared to the reduced cross section $\sigma_{r}$ from
the combined low $Q^{2}$ data [41] and also HERA combined data
[50]. H1 data accompanied with total errors. H1 data  represented
for $Q^{2}=2~\mathrm{GeV}^{2}$ as the closed circles are SVX data,
and the open circles are NVX-BST data, for
$Q^{2}=5~\mathrm{GeV}^{2}$ the closed circles are NVX-BST data and
the open circles are NVX-S9 data, and also for
$Q^{2}=12~\mathrm{GeV}^{2}$ the closed circles are NVX-BST data
[41]. The error bands are due  to the effective parameters in the
parameterization of $F_{2}(x,Q^{2})$ [25] and also the HT
coefficient errors [36-40,42,43]. This comparison with HERA
combined data [50] at $Q^{2}=2$ and $12~\mathrm{GeV}^{2}$ is
defined. }\label{Fig1}
\end{figure}
\newpage{
\begin{table}[h]
\centering \caption{The reduced cross section $\sigma_{r}$
determined based on the nonlinear and higher twist effects in
$Q^{2}$ values $2$, $5$ and $12~\mathrm{GeV}^{2}$ at $x<0.01$.
These results accompanied with the uncertainties due to the
coefficient functions errors [25] and compared with the H1 data
[41] as the uncertainties are quoted in $\%$ relative to
$\sigma_{r}$. }\label{table:table1}
\begin{minipage}{\linewidth}
\renewcommand{\thefootnote}{\thempfootnote}
\centering
\begin{tabular}{|l||c|c||c|c||} \hline\noalign{\smallskip} $Q^{2}(\mathrm{GeV}^{2})$ & $ x$
& $\mathrm{H1}~\mathrm{data}$ &
$ \sigma_{r}{\pm}\delta \%$ & $ \sigma_{r}{\pm}\delta$  \\
\hline\noalign{\smallskip}
2& 2.470E-5& $\mathrm{NVX-S9}~ \mathrm{data}$ & $0.756{\pm}9.23\%$ & $0.833^{+0.165}_{-0.164}$\\
2& 2.928E-5& $\mathrm{SVX}~ \mathrm{data}$ & $0.822{\pm}4.28\%$ & $0.855^{+0.161}_{-0.161}$\\
2& 2.928E-5& $\mathrm{NVX-BST}~ \mathrm{data}$ & $0.788{\pm}4.45\%$ & $0.855^{+0.161}_{-0.161}$\\
2& 5.000E-5& $\mathrm{SVX}~ \mathrm{data}$ & $0.837{\pm}3.10\%$ & $0.847^{+0.152}_{-0.151}$  \\
2& 5.000E-5& $\mathrm{NVX-BST}~ \mathrm{data}$ & $0.792{\pm}5.31\%$ & $0.847^{+0.152}_{-0.151}$  \\
2& 8.000E-5& $\mathrm{SVX}~ \mathrm{data}$ & $0.791{\pm}3.03\%$ & $0.800^{+0.144}_{-0.143}$  \\
2& 1.300E-4& $\mathrm{SVX}~ \mathrm{data}$ & $0.731{\pm}3.28\%$ & $0.742^{+0.136}_{-0.135}$  \\
2& 2.000E-4& $\mathrm{SVX}~ \mathrm{data}$ & $0.700{\pm}3.58\%$ & $0.691^{+0.129}_{-0.129}$  \\
2& 3.200E-4& $\mathrm{SVX}~ \mathrm{data}$ & $0.578{\pm}4.39\%$ & $0.637^{+0.121}_{-0.121}$  \\
2& 3.200E-4& $\mathrm{NVX-BST}~ \mathrm{data}$ & $0.645{\pm}12.2\%$ & $0.637^{+0.121}_{-0.121}$  \\
2& 5.000E-4& $\mathrm{SVX}~ \mathrm{data}$ & $0.528{\pm}3.95\%$ & $0.590^{+0.116}_{-0.115}$  \\
2& 1.000E-3& $\mathrm{SVX}~ \mathrm{data}$ & $0.490{\pm}3.79\%$ & $0.523^{+0.104}_{-0.104}$  \\
2& 1.000E-3& $\mathrm{NVX-BST}~ \mathrm{data}$ & $0.527{\pm}5.93\%$ & $0.523^{+0.104}_{-0.104}$  \\
2& 3.200E-3& $\mathrm{SVX}~ \mathrm{data}$ & $0.424{\pm}4.65\%$ & $0.425^{+0.087}_{-0.087}$  \\
2& 3.200E-3& $\mathrm{NVX-BST}~ \mathrm{data}$ & $0.426{\pm}5.80\%$ & $0.425^{+0.087}_{-0.087}$  \\
5& 6.176E-5& $\mathrm{NVX-S9}~ \mathrm{data}$ & $0.933{\pm}7.70\%$ & $1.002^{+0.233}_{-0.244}$  \\
5& 7.320E-5& $\mathrm{NVX-BST}~ \mathrm{data}$ & $1.052{\pm}3.26\%$ & $1.055^{+0.228}_{-0.240}$  \\
5& 1.300E-4& $\mathrm{NVX-BST}~ \mathrm{data}$ & $1.066{\pm}2.72\%$ & $1.062^{+0.212}_{-0.223}$  \\
5& 2.000E-4& $\mathrm{NVX-BST}~ \mathrm{data}$ & $1.009{\pm}2.62\%$ & $0.997^{+0.201}_{-0.212}$  \\
5& 3.200E-4& $\mathrm{NVX-BST}~ \mathrm{data}$ & $0.911{\pm}2.79\%$ & $0.913^{+0.190}_{-0.199}$  \\
5& 5.000E-4& $\mathrm{NVX-BST}~ \mathrm{data}$ & $0.838{\pm}3.11\%$ & $0.834^{+0.178}_{-0.187}$  \\
5& 8.000E-4& $\mathrm{NVX-BST}~ \mathrm{data}$ & $0.775{\pm}3.50\%$ & $0.754^{+0.166}_{-0.175}$  \\
5& 1.300E-3& $\mathrm{NVX-BST}~ \mathrm{data}$ & $0.686{\pm}2.91\%$ & $0.677^{+0.154}_{-0.162}$  \\
5& 2.000E-3& $\mathrm{NVX-BST}~ \mathrm{data}$ & $0.636{\pm}2.84\%$ & $0.615^{+0.143}_{-0.151}$  \\
5& 3.980E-3& $\mathrm{NVX-BST}~ \mathrm{data}$ & $0.569{\pm}3.18\%$ & $0.523^{+0.125}_{-0.133}$  \\
12& 8.000E-4& $\mathrm{NVX-BST}~ \mathrm{data}$ & $1.067{\pm}3.05\%$ & $1.014^{+0.263}_{-0.284}$  \\
12& 1.300E-3& $\mathrm{NVX-BST}~ \mathrm{data}$ & $0.938{\pm}3.31\%$ & $0.898^{+0.243}_{-0.262}$  \\
12& 2.000E-3& $\mathrm{NVX-BST}~ \mathrm{data}$ & $0.850{\pm}3.00\%$ & $0.802^{+0.224}_{-0.243}$  \\
12& 3.200E-3& $\mathrm{NVX-BST}~ \mathrm{data}$ & $0.752{\pm}2.98\%$ & $0.705^{+0.205}_{-0.222}$  \\
12& 6.310E-3& $\mathrm{NVX-BST}~ \mathrm{data}$ & $0.650{\pm}2.89\%$ & $0.578^{+0.174}_{-0.190}$  \\
\hline\noalign{\smallskip}
\end{tabular}
\end{minipage}
\end{table}
}
\begin{table}[h]
\centering \caption{The values of $\chi^{2}/N$ for the computed
$\sigma_{r}$ to the H1 data [41] in the small and moderate-$Q^{2}$
regions for $x<0.01$ are determined. Also the number of data
points in each case is mentioned. }\label{table:table1}
\begin{minipage}{\linewidth}
\renewcommand{\thefootnote}{\thempfootnote}
\centering
\begin{tabular}{|l||c||c|c||} \hline\noalign{\smallskip} $Q^{2}(\mathrm{GeV}^{2})$
& $\mathrm{H1}~\mathrm{data}$ & N & $ \chi^{2}/N$  \\
\hline\noalign{\smallskip}
& $\mathrm{NVX-S9}~ \mathrm{data}$&  & \\
2& $\mathrm{NVX-BST}~ \mathrm{data}$& 15 & 1.696\\
& $\mathrm{SVX}~ \mathrm{data}$&  & \\
\hline\noalign{\smallskip}
& $\mathrm{NVX-S9}~ \mathrm{data}$&  & \\
5& $\mathrm{NVX-BST}~ \mathrm{data}$& 10 & 1.021\\
\hline\noalign{\smallskip}
12& $\mathrm{NVX-BST}~ \mathrm{data}$& 5 & 5.315\\
\hline\noalign{\smallskip}
\end{tabular}
\end{minipage}
\end{table}
\end{document}